\documentclass[aps,prl,10pt,twocolumn,superscriptaddress]{revtex4}
\usepackage{mathrsfs, amssymb, amsmath}  
\usepackage{footmisc}
\usepackage{latexsym}
\usepackage{natbib, comment}
\usepackage{url}
\usepackage{dcolumn}
\usepackage{multirow}
\usepackage{color}
\usepackage{soul}
\usepackage[normalem]{ulem}
\usepackage{amsfonts,amssymb,amsmath, txfonts}
\usepackage{graphicx,epsfig, cancel}
\usepackage{psfrag}
\usepackage{hyperref}
\hypersetup{colorlinks=true}
\usepackage{mathtools}
\usepackage{enumitem}
\usepackage{float}

\usepackage{epsfig}
\usepackage{amssymb}
\usepackage{amsfonts}
\usepackage{amsmath}
\usepackage{euscript}
\usepackage{verbatim}
\usepackage{latexsym}
\usepackage{graphicx}
\usepackage{caption, color}

\usepackage{subcaption}

\usepackage[all]{xy}

\usepackage[dvipsnames]{xcolor}
\usepackage{xcolor}
\hypersetup{ linktoc=all,
    colorlinks, linkcolor={brightpink},
    citecolor={blue}, urlcolor={blue}
}

\definecolor{rosy}{RGB}{230,235,252}
\definecolor{myframetitle}{RGB}{90,89,170}
\definecolor{myblocktitle}{RGB}{140,185,249}
\definecolor{mytitle}{RGB}{10,80,26}

\definecolor{darkgreen}{RGB}{27,130,45}
\definecolor{darkblue}{rgb}{0,0,0.3}
\definecolor{darkred}{rgb}{0.7,0,0}

\definecolor{light gray}{RGB}{220,220,220}
\definecolor{dark purple}{RGB}{108,0,217}
\definecolor{pink}{RGB}{190,20,100}
\definecolor{orang}{RGB}{193,63,0}
\definecolor{green}{RGB}{11,98,17}
\definecolor{darkpink}{RGB}{153,0,76}
\definecolor{bluegreen}{RGB}{0,102,102}
\definecolor{greenlagan}{RGB}{0,102,0}
\definecolor{redgreen}{RGB}{102,102,0}
\definecolor{Redgreen}{RGB}{153,76,0}
\definecolor{vividviolet}{rgb}{0.62, 0.0, 1.0}
\definecolor{amaranth}{rgb}{0.9, 0.17, 0.31}
\definecolor{palatinateblue}{rgb}{0.15, 0.23, 0.89}
\definecolor{brightpink}{rgb}{1.0, 0.0, 0.5}
\definecolor{cornflowerblue}{rgb}{0.39, 0.58, 0.93}
\definecolor{deepcarminepink}{rgb}{0.94, 0.19, 0.22}
\definecolor{radicalred}{rgb}{1.0, 0.21, 0.37}

\usepackage{listings}

\newcommand{\beq}{\begin{equation}}
\newcommand{\eeq}{\end{equation}}
\newcommand{\bea}{\begin{eqnarray}}
\newcommand{\eea}{\end{eqnarray}}
\newcommand{\beas}{\begin{eqnarray*}}
\newcommand{\eeas}{\end{eqnarray*}}

\newcommand{\bquo}{\begin{quote}}
\newcommand{\enqu}{\end{quote}}
\renewcommand{\(}{\begin{equation}}
\renewcommand{\)}{\end{equation}}

\def\H{ \hbox{\rm H}}

\def\H0{{\text{H}\hspace*{-2.05mm}\text{H} 0\hspace*{-1.35mm}0\ }}

\def\lcdm{$\Lambda$CDM}
\def\be{\begin{equation}}
\def\ee{\end{equation}}
\def\beq{\begin{equation}}
\def\eeq{\end{equation}}
\def\bea{\begin{eqnarray}}
\def\eea{\end{eqnarray}}

\begin{document}

\title{A Tilt Instability in the Cosmological Principle}


\author{Chethan Krishnan}\email{chethan.krishnan@gmail.com}
\affiliation{Center for High Energy Physics,\\
Indian Institute of Science, Bangalore 560012, India}

\author{Ranjini Mondol}\email{ranjinim@iisc.ac.in }
\affiliation{Center for High Energy Physics,\\
Indian Institute of Science, Bangalore 560012, India}

\author{M. M. Sheikh-Jabbari}\email{jabbari@theory.ipm.ac.ir}
\affiliation{School of Physics, Institute for Research in Fundamental Sciences (IPM),\\ P.O.Box 19395-5531, Tehran, Iran}

\begin{abstract}
We show that the Friedmann-Lema\^{i}tre-Robertson-Walker (FLRW) framework has an instability towards the growth of fluid flow anisotropies, even if the Universe is accelerating. This flow (tilt) instability in the matter sector is invisible to Cosmic No-Hair Theorem-like arguments, which typically only flag shear anisotropies in the metric. We illustrate our claims in the setting of ``dipole cosmology'', the maximally Copernican generalization of FLRW that can accommodate a flow. Simple models are sufficient to show that the cosmic flow need not track the  shear, even in the presence of a positive cosmological constant.  We also emphasize that the growth of the tilt hair is fairly generic if the total equation of state $w(t) \rightarrow -1$ at late times (as it does in standard cosmology), irrespective of the precise model of dark energy. {The generality of our theoretical result puts various recent observational claims about late time anisotropies and cosmic dipoles in a new light.}

\end{abstract}

\maketitle

\setcounter{footnote}{0}


\section{Introduction and motivation }

Copernicus put forward the viewpoint that we are not privileged observers in the Universe \cite{Copernicus}. Copernican viewpoint has influenced physics and in particular cosmology ever since \cite{WeinbergOldBook, Ellis-Maartens-McCallum--Book}. After Hubble's discovery in 1929 \cite{Hubble-1929},  the Copernican principle has been dubbed  the ``cosmological principle''. It states that  Universe is homogeneous and isotropic on constant cosmic time slices and is formulated within the FLRW cosmology. The current concordance flat $\Lambda$ Cold Dark Matter (\lcdm) model, sometimes also called the standard model of cosmology, is a specific model within the FLRW framework. 

Flat \lcdm\ model culminates the cosmology today and  it rests upon cosmological data at various different redshifts. However, steadily improving precision in observations has created tensions within the model \cite{crisis, Intro0,Eleonora-et-al-review-1}. Many models, almost all within the FLRW framework, have been proposed to resolve the tensions, but none of them seem fully satisfactory \cite{Perivolaropoulos:2021jda, SNOWMASS-2022}. 
This has promoted the idea  that the current cosmological tensions may be symptoms of a much deeper issue, a violation of the cosmological principle \cite{Upper-bound-H0, binning-fitting}. This viewpoint resonates with the growing concern that the cosmological principle may not pass all observational tests, see \cite{Beyond-FLRW-review} for a recent review.

The simplest (ie., the most Copernican) setting that can accommodate a cosmic flow is the ``dipole cosmology'' paradigm of \cite{KMS} that generalizes the FLRW framework. This is a tilted axially symmetric Bianchi V/VII$_h$ cosmology that falls into the broader rubric of tilted homogeneous models \cite{King, Ellis-lectures, Ellis-Maartens-McCallum--Book}. The ``tilt'' refers to a flow in the fluid that is {\em not} orthogonal to the homogeneous time slices. As such, dipole cosmology is a minimal setting within the tilted models for formulating a cosmic dipole. It has 2 functions of the time coordinate in the background metric -- an overall Hubble expansion rate and the metric anisotropy parametrized by shear. It also has 3 functions in the cosmic fluid sector -- energy density, pressure and the tilt. Part of our goal in this letter is to emphasize that this is an extremely simple and tractable framework that can accommodate a dipole flow, that generalizes the Friedmann equations while still being ODEs. {The simplicity of this set up as a model-building paradigm, especially with non-trivial equations of state including mixtures \cite{Dipole-LCDM}, does not seem to have been emphasized before.}

A second  motivation is that this class of models allow us to investigate certain stability properties of the (conventional) FLRW framework that is often not emphasized. Standard lore dictates that cosmic acceleration washes away the anisotropies. Statements of this kind are usually called Cosmic No-Hair Theorems \cite{Wald-cosmic-no-hair}, and they usually refer to {very fast (exponential) falloff of} the shear anisotropy in the metric. Dipole cosmology enables us to see that even while the shear anisotropies die down, the tilt anisotropies of an accelerating Universe need not. We will illustrate this here in the setting of fluids with constant equations of state together with a positive cosmological constant $\Lambda$. {This is an extremely simple scenario that is implicit in the older work of \cite{Coley-2006}. Our goal is to point out that this is the simplest instantiation of the more general scenario noted in \cite{KMS} where large classes of models with the total equation of state $w(t) \rightarrow -1$ at late times, were shown to have growing tilt anisotropies at late times.}



Let us list a few of our main results. (1) Even when the  shear dies off at late times, the tilt or bulk flow can be relatively large or can even grow {in very large classes of models}. (2) Even with a very small initial value for tilt and almost FLRW background, we can get a sizable bulk flow, {signalling a tilt instability in the FLRW cosmology.} (3) These features can arise even in an accelerating Universe (say, one with a positive cosmological constant). (4) Together with the cases considered in \cite{KMS}, the results of this paper show that these observations, while not absolutely generic, are quite easily realized (and generic indeed, in some regions of parameter/initial-condition space).  The punchline is that at late times, even if we are dealing with metrics which are homogeneous and almost isotropic, we still could have substantial bulk flows and cosmic dipoles. It seems likely that this is of significance for late time observational cosmology and  cosmic tensions, even though to make a precise statement, we will need more detailed model building within the dipole cosmology paradigm, {which we will present in an upcoming paper \cite{Dipole-LCDM}.}

Let us emphasize a simple but potentially confusing point. Dipole (tilt) anisotropy is an observer-dependent concept; even in a homogeneous nearly isotropic metric, non-trivial cosmic flows can result in perceived anisotropies/dipoles by standard observers living on those flows.  A detailed discussion of dipole cosmography will be presented elsewhere \cite{Dipole-Observables}.

Our analysis is done using the dipole cosmology equations (presented below) that generalize the FLRW Friedmann equations. They can be viewed as keeping track of the non-linear evolution of tilt perturbations around the FLRW system. Interpreted this way, our result is a demonstration that FLRW cosmology is not stable against homogeneous, anisotropic tilt perturbations even when the Universe is accelerating. In this work we focus on the main results; more detailed analyses and discussions may be found in previous and upcoming works, in particular in \cite{KMS, Dipole-LCDM, Dipole-Observables}.

\section{Dipole Cosmology, the basic setup}\label{sec:2-basic-setup}

{In this section, we review the dipole cosmology set up \cite{King, KMS}.} The most general cosmological metric in 4 dimensions has 6 functions of space and time. Assuming a cosmic (comoving) time coordinate $t$, and spatial homogeneity on constant time slices, we remain with 3 functions of $t$. In our dipole cosmology besides homogeneity we also assume axisymmetry which removes one more function and hence the metric involves only 2 functions of $t$. The metric in the dipole cosmology setting takes the form 
\bea\label{DipoleMetric}
    ds^{2} = -dt^{2} + a^{2}(t)\left[ e^{4b(t)} dz^{2} +  e^{-2b(t)-2A_{0} z}\big(dx^{2}+dy^{2}\big) \right],
\eea
where $a(t)$ is the over all scale factor, $b(t)$ parameterizes the anisotropy and $A_0\neq 0$ is a constant of dimension of inverse length. While it may be set to 1 by a choice of units, we keep it for later convenience. See \cite{Ellis-Maartens-McCallum--Book, MacCallum, King, Stewart-Ellis, Ellis-lectures, Coley-2006,Ellis-Goliath} for background, and \cite{KMS, upcoming-1} for discussions. We may define the Hubble expansion rate $H(t)$ and the cosmic shear $\sigma (t)$ as usual
\begin{equation}\label{H-sigma}
H:=\frac{\dot{a}}{a},\qquad \sigma:=3\dot{b} 
\end{equation}
where \emph{dot} denotes derivative w.r.t. $t$. When $\sigma=0$ the metric reduces to an open FLRW universe. 

In this metric the $z$ direction is chosen to be along the cosmic flow (the tilt). To see this, we note that the in $(t,z, x,y)$ frame the energy momentum tensor of a perfect fluid with energy density $\rho$ and pressure $p$ is 
\begin{equation}\label{T-tilted}
\begin{split}
&T^{a}{}_{b} = T_{\text{iso}}{}^{a}{}_{b} + T_{\text{tilt}}{}^{a}{}_{b}\\
&T_{\text{iso}}{}^{a}{}_{b}=\text{diag}(-\rho, p,p,p)\\
&T_{\text{tilt}}{}^{a}{}_{b}=(\rho+p)\sinh\beta\left(\begin{array}{cccc}
 -\sinh\beta &  ae^{2b}  \cosh \beta & 0 & 0 \\
-\cosh \beta/(ae^{2b}) &  \sinh\beta  & 0 & 0 \\
 0 & 0 & 0 & 0 \\
 0 & 0 & 0 & 0\\
\end{array}
\right)
\end{split}
\end{equation}
where $T_{\text{iso}}{}^{a}{}_b$ is the energy momentum tensor of a usual isotropic perfect fluid. The off-diagonal terms in $T^a{}_b$ are a manifestation of the non-zero bulk flow along $z$ direction; $\beta$ parameterizes the tilt, the bulk flow. The important point in \eqref{T-tilted} is that when $\rho+p=0$, i.e. for a cosmological constant, $\beta$ drops out and we recover the usual FLRW setup. In other words, a cosmological constant can be viewed as having any value of the tilt.

\section{Evolution equations of dipole cosmology}

In our dipole cosmology we have 5 functions of $t$, $\rho(t), p(t), a(t), b(t), \beta(t)$ which are to be specified by Einstein's field equations
\bea\label{cosmoC}
R_{ab}-\frac12 Rg_{ab}= T_{ab}, 
\eea
or the continuity equations $\nabla^a T_{ab}=0$, yielding
\begin{subequations}\label{EoM-H-sigma}
\begin{align}
H^2-\frac19\sigma^2-\frac{A_0^2}{a^2} e^{-4b}=\frac{\rho}{3}&+\frac13 (\rho+p)\sinh^2\beta
\label{EoM-H-sigma-c}\\
\dot{\sigma}+\sigma \big(3H -\frac{2A_0}{a}&\tanh\beta e^{-2b}\big)=0\label{EoM-sigma}\\
\dot{\rho}+3H(\rho+p)=-(\rho+p&)\tanh\beta(\dot{\beta}-\frac{2A_0}{a} e^{-2b}) \label{Con1-1} \\
\dot{p}+H(\rho+p)= -(\rho+p&)\left( \frac23\sigma+\dot{\beta}\coth{\beta}\right). \label{Con2-1}
\end{align}
\end{subequations}
Moreover, we note that the above equations imply
\begin{equation}\label{EoM-H-sigma-d}\begin{split}
 \sigma &=\frac{1}{4A_0} a e^{2b}(\rho+p)\sinh2\beta
\end{split}
\end{equation}
Note also that for $p=-\rho$, $\beta$ drops out of equations and the solution to the above equations for $\rho>0$ is a de Sitter geometry in an open Universe slicing. 

Among the above 4 equations,  \eqref{EoM-H-sigma-c} and \eqref{Con1-1} are generalization of  the 2 Friedmann equations of the usual FLRW cosmology, while \eqref{EoM-sigma} (or \eqref{EoM-H-sigma-d}) and \eqref{Con2-1} are new and govern dynamics of the shear and the tilt. To solve the above equations one needs to supplement them with another equation, which may be taken to be the equation of state $w(t):=d p/d\rho$.

\section{Examples of dipole cosmology models}\label{sec:DC-evolution}

In a previous paper \cite{KMS}, we considered scenarios where the $w(t)$ mentioned above had the feature that $w(t) \rightarrow -1$ at late times, so that the late Universe accelerates. Many examples of this type were considered and a general criterion was provided for when such models will lead to increasing tilt at late times. In particular, if the approach of $w(t)$ to $-1$ at late times is exponential and sufficiently fast, it was noted that the tilt would grow. It should be emphasized that this is $not$ a particularly stringent demand: indeed, it was pointed out that the effective (time-dependent) equation of state of the standard flat $\Lambda$CDM cosmology satisfies the demand! This indicates that the phenomenon of tilt growth is fairly generic  in the space of physically interesting models. 


In this paper, we will discuss a simple class of accelerating models that also exhibit late time tilt growth. These are models with a fluid with constant equation of state $w$ and  a cosmological constant. {This is a setting that has been studied previously, and our goal is to show that these old results are a specific realization of our broader claims in \cite{KMS}. Our secondary goal is to present the relevant equations in a form that is a transparent generalization of the Freedman equations.} We will see that in these dipole $w$-$\Lambda$ models, tilt can increase at late times if $w$ is stiff enough. This {is consistent with our claim that} tilt growth can be a fairly generic phenomenon in accelerating Universes.

These $w$-$\Lambda$ models, as well as the general class of $w(t) \rightarrow -1$ models of \cite{KMS}, are presented to illustrate that our claims are fairly generically true. The specific models themselves are not {to be taken as realistic}.

\subsection{Dipole \texorpdfstring{$w$-$\Lambda$}{wLambda} models
}\label{sec:dipole-const-w-Eos}

\begin{figure}[htp]
    \centering
        \subfloat[\label{fig:wLambda-a}]{{\includegraphics[width=8.0 cm]{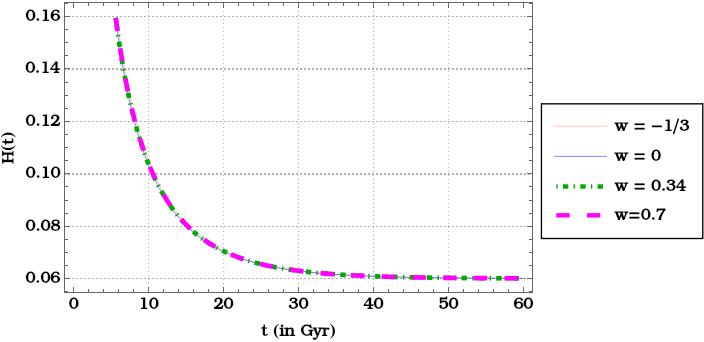}}}\\
   \subfloat[\label{fig:wLambda-sigma}]{{\includegraphics[width=8.5 cm]{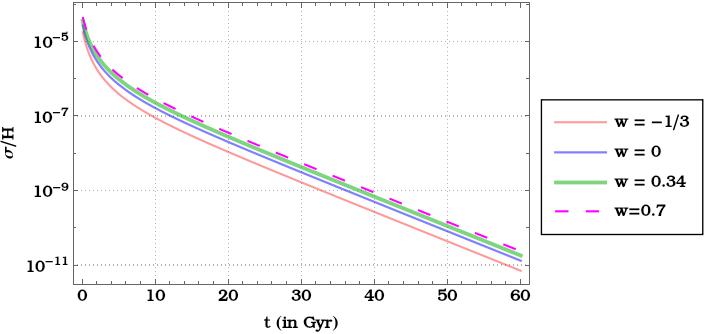} }} \\ \subfloat[\label{fig:wLambda-beta}]{{\includegraphics[width=8.5 cm]{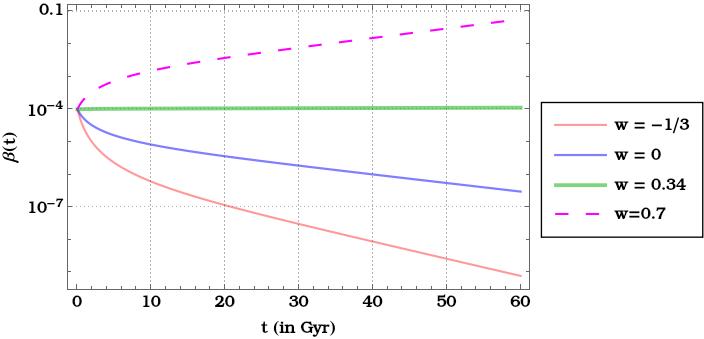} }}  \\ 
 \subfloat[\label{fig:wLambda-beta-growth}]{{\includegraphics[width=8.5 cm]{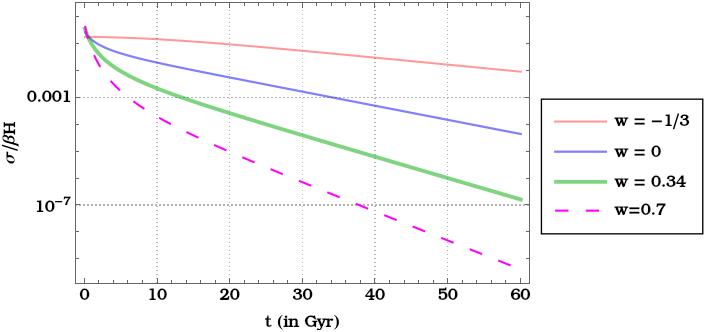} }}     
\caption{Evolution of overall Hubble parameter $H(t)$,  $\sigma/H$, tilt $\beta$ and the dimensionless ratio $\sigma/({\beta}H)$ for dipole $w$-$\Lambda$ model. Initial values are  $a_{in} = 1, b_{in} = 0$, $\rho_{in} = 0.6$, $\beta_{in} = 10^{-4}$ and $\Lambda= 0.0109$. In these plots we have adopted the units in which  $A_0=1$. We have plotted 4 representative values of $w$;   $\beta$ grows for $w>1/3$ at late times. While all the 4 cases have essentially the same $H(t)$ they differ in evolution of $\beta$.}
    \label{fig:w-Lambda-dipole-model}
\end{figure}

As an illustrative example  we study models involving a fluid of constant EoS $w$ and a cosmological constant $\Lambda$. It follows from our earlier discussion that a $\Lambda$ can be incorporated into \eqref{EoM-H-sigma} by defining 
\be\label{w-Lambda-p-rho}
p=w\tilde{\rho}-\Lambda,\qquad \rho=\tilde\rho+\Lambda, \qquad -1< w\leq 1.
\ee
For a generic $w$, \eqref{EoM-H-sigma} implies
\begin{subequations}\label{const-w-dipole-1}
\begin{align}
&\hspace{-7mm}\frac{\ddot{a}}{a}=   \dot{H}+H^2=\frac{\Lambda}{3}-\frac{\tilde\rho}{6}(1+3w)-\frac29 \sigma^2-\frac{\tilde\rho}{3}(1+w)\sinh^2\beta\label{cosmic-acceleration-const-w}\\
&\hspace{-10mm}\dot{\beta}\big(\coth\beta-w \tanh \beta\big)=(3w-1)H-\frac23\sigma-\frac{2w A_0}{a(t)} e^{-2b}\tanh\beta \label{beta-growth-w-const}
\\
&\tilde\rho^{\frac{w}{1+w}} a e^{2b} \sinh\beta = C= const. \label{const-w-dipole-rho-X-beta}\\
&\sigma=\frac{C(1+w)}{2A_0}\tilde\rho^{\frac{1}{1+w}}\cosh\beta \label{sigma-w-dipole}
\end{align}
\end{subequations}
In our analysis we consider non-negative cosmological constant $\Lambda\geq 0$ {and take $\beta>0$}. From these equations we learn:
\begin{enumerate}
    \item As in the FLRW case, for $w\leq -1/3$ we get an accelerated expansion for any $\Lambda\geq 0$.
\item  As Universe expands $a(t)$ grows  and $\rho(t)$ drops.
\item The shear $\sigma(t)$ goes to zero (exponentially fast for accelerated expansion) and the universe isotropizes rapidly. 
\item Since $-1<w\leq 1$, {and for $\beta>0$}, $\coth\beta>1$ and $\tanh\beta <1$, the coefficient of $\dot\beta$ term is always positive. While the last term in \eqref{beta-growth-w-const} does not have a definite sign, it becomes insignificant at late times due to the expansion. Therefore,  for $w>1/3$ the sign of  $\dot\beta$  is positive and the tilt can grow. {This is in accord with the results of \cite{Coley-2006}}. 
\item Cosmological constant $\Lambda$ does not explicitly appear in \eqref{beta-growth-w-const} and therefore it is plausible that $\beta$ growth only depends on  $w$ ($>1/3$) and not the value of $\Lambda$; $\beta$ growth can happen in accelerating/decelerating cosmologies. This is indeed what we have verified numerically. In fact, we find 
that $\beta$ growth is faster in cases with $\Lambda$, {due to faster growth of $H$}.
\end{enumerate}
The above is also confirmed by  numerical evolution of the  equations for $\Lambda=0$ and $\Lambda>0$ cases. We plot the latter  
in FIG.~\ref{fig:w-Lambda-dipole-model}. 

\section{Discussion and outlook}

We have formulated and analysed dipole cosmology, which is constructed to accommodate a cosmic bulk flow in a minimal generalization of the FLRW framework. Our theoretical analyses is based upon Einstein field equations while allowing for a tilt in the energy momentum tensor of the cosmic fluid. It reveals a few important facts. The tilt can remain large and even grow in time, while geometry isotropises (cosmic shear dying off fast), illustrating that FLRW cosmology is unstable against tilt perturbations. This claim is true even in an accelerating Universe, in particular one containing a positive cosmological constant. It is in fact quite generic in Universes where $w(t) \rightarrow -1$ at late times. {This statement is our most significant claim. We also showed that some previous observations about Bianchi models are specific realizations of this claim.} {We also uncovered another interesting feature in some classes of dipole cosmology models \cite{KMS}: while $\beta$ goes to zero at very late times, there could be intermediate stages where the tilt can grow}. While these results are not in the context of a realistic model, they {exhibit the theoretical possibility of instability in FLRW setting due to dipole deformations/perturbations.} {Given the various claims about late time flows and dipole anisotropies in the recent literature, we feel that this observation is worthy of broader notice.} 


Our findings may seem at odds with the common lore, which is based on intuitions gained from Wald's celebrated cosmic no-hair theorem \cite{Wald-cosmic-no-hair} and the fact that inflation leaves us in a Universe in which shear is suppressed. The cosmic evolution afterward does not provide sources for the shear and it remains small in the course of cosmic history. Our analysis emphasizes that tilt and shear are two separate concepts, and our dynamical equations \eqref{EoM-H-sigma} reveal that $\sigma\ll 1$ does not imply $\beta\ll 1$ and in fact $\beta$ can be or become of order 1. These results are compatible with the usual lore, but they highlight the important distinction between having an (almost) isotropic and homogeneous metric and having a homogeneous and isotropic cosmology, which is usually taken to be synonymous. In particular, an observer living in a flowing fluid of galaxies will see a homogeneous Universe with a dipole anisotropy even if the shear ansiotropies in the metric are small \cite{Dipole-Observables}. Viewed in this light, the  CMB observations which are usually taken as observational confirmation of the usual FLRW cosmologies (modulo CMB anomalies \cite{Beyond-FLRW-review, CMB-anomalies}), primarily indicate isotropy of the background metric over which the light propagates and does not exclude bulk flows and cosmic dipoles \cite{Dipole-Observables}. 

Our analysis prompts several theoretical  and observational questions, the most important of which being instability of the homogeneous and isotropic cosmologies to \textit{homogeneous} tilt perturbations; even if one starts with an isotropic Universe after inflation, {a small tilt perturbation in the course of cosmic evolution can yield a sizable bulk flow or cosmic dipole}. It may be interesting to consider models with mixtures of flowing fluids as a setting for model building in dipole cosmology. This will provide a very simple model building paradigm that generalizes the FLRW setting. In the present work, we loosely attributed the tilt $\beta(t)$ to bulk flows and to cosmic dipoles. But in order to properly connect to observations, one should make this attribution precise by developing cosmography in the dipole cosmology setting. At the observational level, it may be interesting to extend the analysis of current and future data to search for traces of a dipole or a bulk flow, see \cite{Beyond-FLRW-review} for a recent review. 


\section*{Acknowledgments}

We thank Ehsan Ebrahimian, Rajeev Jain, Roya Mohayaee, Eoin O Colgain, Mohamed Rameez and Subir Sarkar for discussions. The work of MMShJ is supported in part by SarAmadan grant No ISEF/M/401332. 




\begin{thebibliography}{99}

\bibitem{Copernicus}
https://en.wikipedia.org/wiki/De\_revolutionibus\_orbium\_coelestium

%

\bibitem{WeinbergOldBook}
S.~Weinberg,
``Gravitation and Cosmology: Principles and Applications of the General Theory of Relativity,'' John Wiley and Sons, 1972;
``Cosmology,'' Oxford University press, 2008.

\bibitem{Ellis-Maartens-McCallum--Book}
Ellis, G. F. R., Maartens, R. and MacCallum, M. A. H. (2012). Relativistic cosmology. Cambridge University Press. https://doi.org/10.1017/CBO9781139014403


\bibitem{Hubble-1929}
E. Hubble, ``A relation between distance and radial velocity among extra-galactic nebulae,'' PNAS USA 15 (1929) 168–173.








  
\bibitem{crisis}
L.~Verde, T.~Treu and A.~G.~Riess,
``Tensions between the Early and the Late Universe,''
Nature Astron. \textbf{3}, 891
[arXiv:1907.10625 [astro-ph.CO]].

\bibitem{Intro0}
E.~Di Valentino, L.~A.~Anchordoqui, O.~Akarsu, Y.~Ali-Haimoud, L.~Amendola, N.~Arendse, M.~Asgari, M.~Ballardini, S.~Basilakos and E.~Battistelli, \textit{et al.}
``Cosmology intertwined II: The Hubble constant tension,''
Astropart. Phys. \textbf{131}, 102605 (2021)
[arXiv:2008.11284 [astro-ph.CO]].

 
\bibitem{Eleonora-et-al-review-1}
E.~Di Valentino, O.~Mena, S.~Pan, L.~Visinelli, W.~Yang, A.~Melchiorri, D.~F.~Mota, A.~G.~Riess and J.~Silk,
``In the realm of the Hubble tension\textemdash{}a review of solutions,''
Class. Quant. Grav. \textbf{38} (2021) no.15, 153001
[arXiv:2103.01183 [astro-ph.CO]].



\bibitem{Perivolaropoulos:2021jda}
L.~Perivolaropoulos and F.~Skara,
``Challenges for $\Lambda$CDM: An update,''
[arXiv:2105.05208 [astro-ph.CO]].


\bibitem{SNOWMASS-2022}
E.~Abdalla, G.~Franco Abell\'an, A.~Aboubrahim, A.~Agnello, O.~Akarsu, Y.~Akrami, G.~Alestas, D.~Aloni, L.~Amendola and L.~A.~Anchordoqui, \textit{et al.}
``Cosmology intertwined: A review of the particle physics, astrophysics, and cosmology associated with the cosmological tensions and anomalies,''
JHEAp \textbf{34} (2022), 49-211
doi:10.1016/j.jheap.2022.04.002
[arXiv:2203.06142 [astro-ph.CO]].




\bibitem{Upper-bound-H0}
C.~Krishnan, R.~Mohayaee, E.~\'O.~Colg\'ain, M.~M.~Sheikh-Jabbari and L.~Yin,
``Does Hubble tension signal a breakdown in FLRW cosmology?,''
Class. Quant. Grav. \textbf{38} (2021) no.18, 184001
[arXiv:2105.09790 [astro-ph.CO]];

C.~Krishnan, R.~Mohayaee, E.~\'O.~Colg\'ain, M.~M.~Sheikh-Jabbari and L.~Yin,
``Hints of FLRW breakdown from supernovae,''
Phys. Rev. D \textbf{105} (2022) no.6, 063514
[arXiv:2106.02532 [astro-ph.CO]].

\bibitem{binning-fitting}

E.~\'O.~Colg\'ain, M.~M.~Sheikh-Jabbari, R.~Solomon, M.~G.~Dainotti and D.~Stojkovic,
``Putting Flat $\Lambda$CDM In The (Redshift) Bin,''
[arXiv:2206.11447 [astro-ph.CO]].

E.~\'O.~Colg\'ain, M.~M.~Sheikh-Jabbari and R.~Solomon,
``High Redshift $\Lambda$CDM Cosmology: To Bin or not to Bin?,''
[arXiv:2211.02129 [astro-ph.CO]].

\bibitem{Beyond-FLRW-review}
P.~K.~Aluri, P.~Cea, P.~Chingangbam, M.~C.~Chu, R.~G.~Clowes, D.~Hutsem\'ekers, J.~P.~Kochappan, A.~Krasi\'nski, A.~M.~Lopez and L.~Liu, \textit{et al.}
``Is the Observable Universe Consistent with the Cosmological Principle?,''
[arXiv:2207.05765 [astro-ph.CO]].


\bibitem{KMS}
C.~Krishnan, R.~Mondol and M.~M.~Sheikh-Jabbari,
``Dipole Cosmology: The Copernican Paradigm Beyond FLRW,''
[arXiv:2209.14918 [astro-ph.CO]].


\bibitem{King}
A.~R.~King and G.~F.~R.~Ellis,
``Tilted homogeneous cosmological models,''
Commun. Math. Phys. \textbf{31} (1973), 209-242

\bibitem{Ellis-lectures}
G.~F.~R.~Ellis and H.~van Elst,
``Cosmological models: Cargese lectures 1998,''
NATO Sci. Ser. C \textbf{541} (1999), 1-116
[arXiv:gr-qc/9812046 [gr-qc]].

\bibitem{Dipole-LCDM}

E. Ebrahimian, C. Krishnan, R. Mondol, M.M. Sheikh-Jabbari, \textit{Towards A Realistic Dipole Cosmology: The Dipole $\Lambda$CDM Model}, \texttt{To appear soon}.


\bibitem{Wald-cosmic-no-hair}
R.~M.~Wald,
``Asymptotic behavior of homogeneous cosmological models in the presence of a positive cosmological constant,''
Phys. Rev. D \textbf{28} (1983), 2118-2120

\bibitem{Coley-2006}
A.~A.~Coley, S.~Hervik and W.~C.~Lim,
``Fluid observers and tilting cosmology,''
Class. Quant. Grav. \textbf{23} (2006), 3573-3591
[arXiv:gr-qc/0605128 [gr-qc]].



\bibitem{Dipole-Observables}

E. Ebrahimian, C. Krishnan, R. Mondol, M.M. Sheikh-Jabbari, \textit{Observables in Dipole Cosmologies}, \texttt{To appear soon}.

\bibitem{upcoming-1}
C.~Krishnan and R.~Mondol, \textit{Work in progress}.


\bibitem{Stewart-Ellis}
J.~M.~Stewart and G.~F.~R.~Ellis,
``Solutions of Einstein's equations for a fluid which exhibit local rotational symmetry,''
J. Math. Phys. \textbf{9} (1968), 1072-1082
doi:10.1063/1.1664679

\bibitem{MacCallum}
G.~F.~R.~Ellis and M.~A.~H.~MacCallum,
``A Class of homogeneous cosmological models,''
Commun. Math. Phys. \textbf{12} (1969), 108-141


\bibitem{Ellis-Goliath}
M.~Goliath and G.~F.~R.~Ellis,
``Homogeneous cosmologies with cosmological constant,''
Phys. Rev. D \textbf{60}, 023502 (1999)
[arXiv:gr-qc/9811068 [gr-qc]].



\bibitem{CMB-anomalies}
D.~J.~Schwarz, C.~J.~Copi, D.~Huterer and G.~D.~Starkman,
``CMB Anomalies after Planck,''
Class. Quant. Grav. \textbf{33} (2016) no.18, 184001
[arXiv:1510.07929 [astro-ph.CO]].





\end{thebibliography}
\end{document}